\documentclass[nonblindrev]{informs3} % current default for manuscript submission

\usepackage{balance}
\usepackage[tight,footnotesize]{subfigure}
\usepackage{graphicx}
\usepackage{algorithm}
\usepackage[noend]{algorithmic}
\usepackage{multirow}
\usepackage{slashbox}
\usepackage{caption}
\OneAndAHalfSpacedXII % current default line spacing
\usepackage{natbib}
 \bibpunct[, ]{(}{)}{,}{a}{}{,}%
 \def\bibfont{\small}%
\TheoremsNumberedThrough     % Preferred (Theorem 1, Lemma 1, Theorem 2)
\EquationsNumberedThrough    % Default: (1), (2), ...
\MANUSCRIPTNO{2019} % When the article is logged in and DOI assigned to it,
\begin{document}
%%%%%%%%%%%%%%%%

%\setcounter{page}{1}
%\MONTH{May}% (month or a similar seasonal id)
%\YEAR{2015}% e.g., 2005
%\SHORTYEAR{15}% shortened year (two-digit)
%\DOI{10.1287/xxxx.0000.0000}%

\RUNAUTHOR{Tang}

\RUNTITLE{Price of Dependence: Stochastic Submodular Maximization with Dependent Items}

% \TITLE{Bundling Infformation Goods of Decreasing Value}
\TITLE{Price of Dependence: Stochastic Submodular Maximization with Dependent Items}

\ARTICLEAUTHORS{%
\AUTHOR{Shaojie Tang}
\AFF{The University of Texas at Dallas}
} % end of the block

\ABSTRACT{In this paper, we study the stochastic submodular maximization problem with dependent items subject to packing constraints such as matroid and knapsack constraints. The input of our problem is a finite set of items, and each item  is in a particular state from a set of possible states. After picking an item, we are able to observe its state. We assume a monotone and  submodular utility function over items and states, and our objective is to select a group of items adaptively so as to maximize the expected utility. Previous studies on stochastic submodular maximization often assume that items' states are independent, however, this assumption may not hold in general. This motivates us to study  the stochastic submodular maximization problem  with dependent items. We first introduce the concept of \emph{degree of independence} to capture the degree to which one item's state is dependent on others'. Then we propose a non-adaptive policy that approximates the optimal
adaptive policy within a factor of  $\alpha(1-e^{-\frac{\kappa}{2}+\frac{\kappa}{18m^2}}-\frac{\kappa+2}{3m\kappa})$ where the value of  $\alpha$ is depending on the type of constraints, e.g., $\alpha=1$ for matroid constraint, $\kappa$ is the degree of independence, e.g., $\kappa=1$ for independent items, and $m$ is the number of items. We also analyze the adaptivity gap, i.e., the ratio of the values of  best adaptive policy and  best non-adaptive policy,  of our problem with prefix-closed constraints.}%Since our policy is non-adaptive, $\alpha(1-e^{-\frac{\kappa}{2}+\frac{\kappa}{18m^2}}-\frac{\kappa+2}{3m\kappa})$ is a valid upper bound on the adaptivity gap, i.e., the ratio of the values of  best adaptive policy and  best non-adaptive policy is bounded by $\alpha(1-e^{-\frac{\kappa}{2}+\frac{\kappa}{18m^2}}-\frac{\kappa+2}{3m\kappa})$ .}

%\KEYWORDS{approximation algorithm; team formation; cover decomposition}

\maketitle
%%%%%%%%%%%%%%%%%%%%%%%%%%%%%%%%%%%%%%%%%%%%%%%%%%%%%%%%%%%%%%%%%%%%%%

% Samples of sectioning (and labeling) in IJOC
% NOTE: (1) \section and \subsection do NOT end with a period
%       (2) \subsubsection and lower need end punctuation
%       (3) capitalization is as shown (title style).
%
%\section{Introduction.}\label{intro} %%1.
%\subsection{Duality and the Classical EOQ Problem.}\label{class-EOQ} %% 1.1.
%\subsection{Outline.}\label{outline1} %% 1.2.
%\subsubsection{Cyclic Schedules for the General Deterministic SMDP.}
%  \label{cyclic-schedules} %% 1.2.1
%\section{Problem Description.}\label{problemdescription} %% 2.

\section{Introduction}\label{sec:introduction}

Stochastic submodular maximization (SSM) and its variants have been extensively studied recently \citep{asadpour2008stochastic}\cite{asadpour2015maximizing}\cite{adamczyk2016submodular}\cite{hellerstein2015discrete}. The input of SSM is a set of items, each item belongs to a particular state from a set of possible states. After picking an item, we are able to observe its state. Given a monotone and submodular utility function over all items and their states, our objective is to adaptively select a group of items that maximize the expected utility subject to a variety of constraints. One example is  stochastic sensor cover problem. In this example, we are given a set of sensors and the state of each sensor is the subset of targets it covers, this subset may change due to uncertain environmental conditions. After selecting a sensor, we are able to observe its state, i.e., the actual subset of targets that can be covered by this sensor. Then the objective of stochastic sensor cover problem is to adaptively select a group of sensors to cover the largest amount of targets (in expectation).

Majority of existing work assume that items are independent from each other, i.e., one item's state does not depend on others'. However, this assumption does not always hold in reality. Consider the example of stochastic maximum $k$-cover, since each sensor's state is affected by the environmental conditions which are shared among all sensors, thus their states are correlated. Another example is from viral marketing \citep{yuan2017discount,yuan2017no} where one customer's influence in the social network is correlated with her neighbors' influence. \cite{golovin2011adaptive} extended the previous studies to dependent items, however, their  results only hold when the utility function is  \emph{adaptive submodular}. It is not clear how to generalize their results to more general settings when adaptive submodularity does not hold. Very recently, \cite{fujii2019beyond} propose a concept named \emph{adaptive submodularity ratio} to study the performance of greedy policy subject to cardinality constraint. However, their analysis  cannot be applied to general constraints.

In this paper, we  study the SSM with dependent items (SSMDI) subject to general constraints such as downward-closed and prefix-closed constraints. To capture the degree to which one item's state is correlated with others', we introduce the concept of \emph{degree of independence}. A larger degree of independence indicates a weaker correlation among all items' states, i.e., this value is 1 for independent items.  Then we propose a non-adaptive policy based on the optimistic continuous greedy algorithm. We say a policy is non-adaptive if it always picks the next item before observing the states of picked items.  We show that our non-adaptive policy achieves approximation ratio $\alpha(1-e^{-\frac{\kappa}{2}+\frac{\kappa}{18m^2}}-\frac{\kappa+2}{3m\kappa})$ where the value of  $\alpha$ is depending on the type of constraints, e.g., $\alpha=1$ for matroid constraint, $\kappa$ is the degree of independence, and $m$ is the number of items. Since our policy is non-adaptive,   $\frac{1}{\alpha(1-e^{-\frac{\kappa}{2}+\frac{\kappa}{18m^2}}-\frac{\kappa+2}{3m\kappa})}$ is also known as an upper bound on  the adaptivity gap, i.e.,  the ratio of the values of best adaptive and best non-adaptive policies is bounded by $\frac{1}{\alpha(1-e^{-\frac{\kappa}{2}+\frac{\kappa}{18m^2}}-\frac{\kappa+2}{3m\kappa})}$.  As a negative result, \cite{asadpour2015maximizing} show that the adaptivity gap with dependent items can be at least $m/2$. At the end of this paper, we also analyze the adaptivity gap of SSMDI with prefix-closed constraints.

\section{Preliminaries}
\subsection{Submodular Function and Multilinear Extension}
\label{sec:pre}
A submodular function is a set function $f:2^{\Omega}\rightarrow \mathbb{R}$, where $2^{\Omega } $ denotes the power set of  $\Omega$, which satisfies a natural ``diminishing returns" property: the marginal gain from adding an element to a set $X$ is at least as high as the marginal gain from adding the same element to a superset of $X$. Formally, a submodular function satisfies the follow property: For every $X, Y \subseteq \Omega$ with $X \subseteq Y$ and every $x \in \Omega \backslash Y$, we have that $f(X\cup \{x\})-f(X)\geq f(Y\cup \{x\})-f(Y)$. We say a submodular function $f$ is monotone if $f(X) \leq f(Y)$ whenever $X \subseteq Y$.  Consider any vector $\bold{x}\in [0,1]^n$. The multilinear extension of $f$ is defined as $F(\bold{x})=\sum_{X\subseteq \Omega} f(X) \prod_{i\in X} x_i \prod_{i\notin X} (1-x_i)$.

\section{Notations and Problem Formulation}
\label{sec:system}
\subsection{Items and States} Let $E$ denote a finite set of $m$ items, and each item $e\in E$ is in a particular state from a set $O$ of possible states.  Let $\phi: E\rightarrow O$ denote a realization of item states.   Let $\Phi=\{\Phi_e|e\in E\}$ be a random realization where $\Phi_e$ denotes a random realization of $e$.   After picking an item $e$, we are able to observe its realization $\Phi_e$.  Let $\mathcal{U}$ denote the set of all realizations, we assume there is a known prior probability distribution $\mathcal{D}$ over realizations, i.e., $\mathcal{D}=\{\Pr[\Phi=\phi]: \phi\in \mathcal{U}\}$. Given a set of items $S$, let $\Phi_S=\{\Phi_e\mid e\in S\}$ denote the random realization of $S$ and  $\mathcal{U}_S$ denote the set of all possible realizations that involve $S$. Let $\mathcal{D}_e(\phi_S)=\{\Pr[\Phi_e=\phi_e\mid \Phi_S=\phi_S]: \phi_e\in \mathcal{U}_e\}$ denote the prior probability distribution over realizations conditioned on $\Phi_S=\phi_S$.%Let $\psi_U: U\rightarrow O $ denote the partial realization observed after picking  $U\subseteq E$.
%For any $Z\subseteq E$, we say a partial realization $\psi_Z$ is feasible if $\Pr[\Psi_Z=\psi_Z]> 0$, i.e., $\psi_Z$ is observed with positive probability. We use $O_Z \subseteq  Z\times O$ to denote the entire set of feasible realizations of $Z$.

\subsection{Utility Function and Problem Formulation} Let $f: 2^{E\times O}\rightarrow \mathbb{R}_{\geq 0}$ be a monotone and submodular function over all items and their states. A policy $\pi$ is a function that specifies which item to pick next under the  observations made so far: $\pi: 2^{E\times O}\rightarrow E$. Note that $\pi$ can be regarded as some decision tree that specifies a rule for  picking items adaptively. Let $\mathcal{I}$ be a  \emph{downward-closed} family of subsets of $E$. Let $E(\pi, \phi)$ denote the subset of items picked by policy $\pi$ under $\phi$.  Then the utility of $\pi$ can be expressed as $f(\pi)= \sum_{\phi \in \mathcal{U}} \beta_\phi f(\cup_{e\in E(\pi, \phi)}\phi_e)$ where $\beta_\phi$ denotes the probability that $\phi$ is realized. We say a policy $\pi$ is \emph{feasible} if for any $\phi$, $E(\pi, \phi)\in \mathcal{I}$. Our goal is to identify the best feasible policy that maximizes its expected utility.
\[\max_{\pi} f(\pi) \mbox{ subject to $E(\pi, \phi)\in \mathcal{I}$ for any $\phi$.}\]

 %Given any partial realizations $\psi_U \in U\times O$ and $\psi_V \in V\times O$ of two disjoint sets of items $U$ and $V$, we use $\Pr[\psi_U|\psi_V]$ to denote the probability of $\psi_U$ being observed conditioned on $\psi_V$ are observed. %In case $\psi_V$ is not a feasible realization (i.e., the probability that $\psi_V$ is observed is zero), we set $\Pr[\psi_U|\psi_V]=0$ for all $\psi_U \in U\times O$.

 \subsection{More Notations and Degree of Independence}

 By abuse of notation, define $f(S)=\mathbb{E}_{\Phi\sim \mathcal{D}}[f(\cup_{v\in S} \Phi_v)]$ as the value of items $S\subseteq E$ where $\Phi\sim \mathcal{D}$ means $\Phi$ follows the distribution of $\mathcal{D}$. Let $f_S(e)=f(S\cup \{e\})-f(S)$ denote the marginal value of item $e$ with respect to $S$. Define $f_S(\phi_e)=\mathbb{E}_{\Phi\sim \mathcal{D}}[f((\cup_{v\in S} \Phi_v)\cup \phi_e )]-f(S)$ as the marginal value of item $e$'s state $\phi_e$ with respect to $S$.

Given a vector $\bold{x}\in [0,1]^m$, let $R$ be a random set obtained by picking each item $e$ independently with probability $x_e$, then the multilinear extension $F(\bold{x})$ is defined as the expected value of  $f(R)$: $F(\bold{x})=\sum_{S\subseteq V} f(S) \prod_{e\in S} x_e \prod_{e\notin R} (1-x_e)$.  Let $F_\mathbf{x}(e)=\mathbb{E}[f_R(e)]$ denote the marginal value of item $e$ with respect to $\mathbf{x}$. Define $F_\mathbf{x}(\phi_{e})=\mathbb{E}[f_R(\phi_{e})]$ as the marginal value of  item $e$'s state $\phi_e$ with respect to $\mathbf{x}$. For notation convenience, denote $\bold{x}\setminus e$ as a new vector by setting $x_e$ in $\bold{x}$ to be 0.

 We next introduce the concept of \emph{degree of independence} which refers to the degree to which one item's state is correlated the others'.
\begin{definition} \label{defin}[Degree of Independence] The degree of independence $\kappa(\mathcal{D})\in[0,1]$ of  a known prior probability distribution $\mathcal{D}$ is defined as follows
\begin{equation}
\kappa(\mathcal{D}):=\min_{  e\in E,  S\subseteq E\setminus \{e\}, V\subseteq E\setminus \{e\}, \phi_{V}\in \mathcal{U}_V} \frac{f_S(e)}{\mathbb{E}_{\Phi_e \sim \mathcal{D}_e(\phi_V)} [f_S(\Phi_e)]}
\end{equation}
 If the numerator and denominator are both 0, the ratio is considered to be 1.
\end{definition}

Notice that if all items' states are realized independently from each other, we have $\forall e\in E, \forall S\subseteq E\setminus \{e\},  \forall V\subseteq E\setminus \{e\}, \forall \phi_{V}\in \mathcal{U}_V: f_S(e)=\mathbb{E}_{\Phi_e \sim \mathcal{D}_e(\phi_V)} [f_S(\Phi_e)]$ due to the state of $e$ does not dependent on other items' states. It follows that $\kappa(\mathcal{D})=1$ when items are independent. We omit $\mathcal{D}$ from $\kappa(\mathcal{D})$  if it is clear from the context.

\section{Algorithm Design}
\label{sec:1}
In this section, we present a \textbf{non-adaptive policy} $\pi^g$, later, we show that the ratio of the utilities of $\pi^g$ to  \textbf{best adaptive policy} is bounded. This result is also known as adaptivity gap. The general idea of $\pi^g$ is to first find a fractional solution using the optimistic  continuous greedy algorithm (Algorithm \ref{alg:greedy-peak}) and then round it to an integral solution.

\paragraph{Stage 1: Optimistic Continuous Greedy Algorithm} We first explain the design of the optimistic continuous greedy algorithm. Algorithm \ref{alg:greedy-peak} maintains a fractional solution $\mathbf{y}(t)$, starting with $\mathbf{y}(0)=(0,0,\cdots,0)$. Let  $R(t)$ be a random set which contains each $e$ independently with probability $y_e(t)$. In each round $t$, it updates the weight $F_{\mathbf{y}(t)\setminus e}(e)$ of each item $e\in E$ as follows,
\begin{equation}
\label{eq:1}
F_{\mathbf{y}(t)\setminus e}(e)=\mathbb{E}[f(R_{\overline{e}}(t)\cup\{e\})]-\mathbb{E}[f(R_{\overline{e}}(t))]
 \end{equation}
 where $R_{\overline{e}}(t)=R(t)\setminus e$ is a subset of $R(t)$ by excluding $e$. As compared with the standard continuous greedy algorithm \citep{calinescu2011maximizing}, we define the weight of each item in a different way, i.e., \cite{calinescu2011maximizing} adopt the following weight function $F_{\mathbf{y}(t)}(e)=\mathbb{E}[f(R(t)\cup\{e\})]-\mathbb{E}[f(R(t))]$. We call our algorithm optimistic continuous greedy algorithm due to $F_{\mathbf{y}(t)\setminus e}(e)\geq F_{\mathbf{y}(t)}(e)$, i.e., the expected marginal contribution of any item $e$ defined in our algorithm is no less than the one defined in the standard continuous greedy algorithm.  Eq. (\ref{eq:1}) is also different from the one defined in \cite{asadpour2015maximizing}. They define the weight of $e$  as $\mathbb{E}[f(\Theta_{R(t)}^{\uparrow e})]-\mathbb{E}[f(\Theta_{R(t)})]$ which only makes sense when the state of each item belongs to $\mathbb{R}_+$ (See Section 2 in \cite{asadpour2015maximizing}).

 Since we are not able to obtain the exact value of $F_{\mathbf{y}(t)\setminus e}(e)$, we estimate this value   by averaging over $\frac{10}{\delta^2}(1+\ln m)$ independent samples of $R_{\overline{e}}(t)$. Let $\widetilde{F}_{\mathbf{y}(t)\setminus e}(e)$ denote the estimated value of $F_{\mathbf{y}(t)\setminus e}(e)$. Assume  $P_\mathcal{I} = \mathrm{conv} \{\mathbf{1}_I: I\in \mathcal{I}\}$, the convex relaxation for $\mathcal{I}$, is a down-monotone solvable polytope, we solve the following optimization problem.

\begin{center}
\framebox[0.45\textwidth][c]{
\enspace
\begin{minipage}[t]{0.45\textwidth}
\small
\textbf{P1:}
\emph{Maximize $\sum_{e\in E}\widetilde{F}_{\mathbf{y}(t)\setminus e}(e) y_e$ }\\
\textbf{subject to:} $\mathbf{y}\in P_\mathcal{I}$
\end{minipage}
}
\end{center}
\vspace{0.1in}

After solving \textbf{P1} at round $t$ and obtain an optimal solution $\overline{\mathbf{y}}(t)$, we update the fractional solution at round $t$ as follows $y_{e}(t+\delta)=y_{e}(t)+\delta\overline{{y}}_{e}(t)$. After $1/\delta$ rounds where $\delta=\frac{1}{9m^2}$,  $\mathbf{y}(1)$ is returned as the final solution.

\paragraph{Stage 2: Rounding Fractional Solution}In the second stage, we round the fractional solution $\mathbf{y}(1)$ to an integral solution. As shown in \cite{chekuri2014submodular}, if there exists a $\alpha$-balanced contention resolution scheme for $\mathcal{I}$ where $\alpha\in [0,1]$, then we can find a feasible solution $I\in \mathcal{I}$ such that $\mathbb{E}[f(I)]\geq \alpha F(\mathbf{y}(1))$. It turns out many useful constraints admit good $\alpha$-balanced contention resolution schemes, including matroid constraints and knapsack constraints. For example, when $\mathcal{I}$ specifies an intersection of $k$ matroids and $O(1)$ knapsacks, we have $\alpha=\frac{0.38}{k}$. Notice that when $\mathcal{I}$ specifies a matroid constraint, we can apply pipage rounding technique \citep{ageev2004pipage} to find a feasible solution with expected utility $F(\mathbf{y}(1))$, i.e., $\alpha=1$.

\begin{algorithm}[h]
{\small
\caption{Optimistic Continuous Greedy}
\label{alg:greedy-peak}
%\textbf{Input:} Social network $\mathcal{G}$, budget $\mathcal{B}$, individual attention constraint $\kappa_i$, overall attention constraint $K$.\\
%\textbf{Output:} Seed set $\mathcal{S}$.
\begin{algorithmic}[1]
\STATE Set $\delta=\frac{1}{9m^2}, t=0, \mathbf{y}(0)=(0,0,\cdots,0)$.
\WHILE{$t<1$}
\STATE Let $R(t)$ be a random set which contains each $e$ independent with probability $y_e (t)$.
\STATE For each $e\in E$, let $R_{\overline{e}}(t)=R(t)\setminus e$, estimate
\[F_{\mathbf{y}(t)\setminus e}(e)=\mathbb{E}[f(R_{\overline{e}}(t)\cup\{e\})]-\mathbb{E}[f(R_{\overline{e}}(t))]\]
\STATE Solve $\textbf{P1}$ with estimated weight $\widetilde{F}_{\mathbf{y}(t)\setminus e}(e) y_e$ for each $e$ and obtain an optimal solution $\overline{\mathbf{y}}(t)$
\STATE
\begin{center}
\framebox[0.9\textwidth][c]{
\enspace
\begin{minipage}[t]{0.9\textwidth}
\small
\textbf{P1:}
\emph{Maximize $\sum_{e\in E}\widetilde{F}_{\mathbf{y}(t)\setminus e}(e) y_e$ }\\
\textbf{subject to:} $\mathbf{y}\in P_\mathcal{I}$ where $P_\mathcal{I} = \mathrm{conv} \{\mathbf{1}_I: I\in \mathcal{I}\}$ is a down-monotone
solvable polytope.
\end{minipage}
}
\end{center}
\vspace{0.1in}
\STATE Let $y_{e}(t+\delta)=y_{e}(t)+\delta\overline{{y}}_{e}(t)$; \label{line:1}
\STATE Increment $t=t+\delta$;
\ENDWHILE
\end{algorithmic}
}
\end{algorithm}

%Let $f_S(e) = f(S \cup \{e\}) - f(S)$ denote the marginal value of item $e$ with respect to $S$.
\section{Performance Analysis}
In this section, we prove the following main results.

\begin{theorem}
Assume $\pi^\diamond$ is the optimal policy and there exists a $\alpha$-balanced contention resolution scheme for $\mathcal{I}$, we have $f(\pi^g)\geq \alpha (1-e^{-\frac{\kappa}{2}+\frac{\kappa}{18m^2}}-\frac{\kappa+2}{3m\kappa})f(\pi^\diamond)$ where $\kappa$ is the degree of independence.
\end{theorem}

\emph{Proof:} Observing that if there exists a $\alpha$-balanced contention resolution scheme for $\mathcal{I}$, we can find a feasible solution $I\in \mathcal{I}$ such that $\mathbb{E}[f(I)]\geq \alpha F(\mathbf{y}(1))$, it suffice to prove that $F(\mathbf{y}(1))\geq (1-e^{-\frac{\kappa}{2}+\frac{\kappa}{18m^2}}-\frac{\kappa+2}{3m\kappa}) f(\pi^\diamond)$. Thus, in the rest of the proof, we focus on proving this inequality.

For every $e\in E$, let $y^\diamond_e$ denote the probability that $e$ is picked by $\pi^\diamond$. Due to $\pi^\diamond$ is a feasible policy, $\mathbf{y}^\diamond$ is a convex combination of feasible solutions in $\mathcal{I}$, thus, $y^\diamond_e \in P_{\mathcal{I}}$. We first prove that $ f(\pi^\diamond)\leq F(\mathbf{x})+  \kappa \sum_{e\in E}y^\diamond_{e}   F_{\mathbf{x}\setminus e}(e)$ for any vector $\mathbf{x}\in [0,1]^m$.
  \begin{eqnarray}
 f(\pi^\diamond)= \sum_{\phi \in \mathcal{U}} \beta_\phi f(\cup_{e\in E(\pi^\diamond, \phi)}\phi_e) &\leq&  \sum_{\phi\in \mathcal{U}} \beta_\phi (F(\mathbf{x}) + \sum_{e\in E(\pi^\diamond, \phi)} F_{\mathbf{x}}(\phi_e))\nonumber\\
  &\leq&  \sum_{\phi\in \mathcal{U}} \beta_\phi (F(\mathbf{x}) + \sum_{e\in E(\pi^\diamond, \phi)} F_{\mathbf{x} \setminus e}(\phi_{e})) \nonumber\\
  &\leq& F(\mathbf{x})+  \sum_{\phi\in \mathcal{U}}\sum_{e\in E(\pi^\diamond, \phi)} (\beta_\phi F_{\mathbf{x}\setminus e}(\phi_{e}))\nonumber\\
  %&\leq& f(H)+  \sum_{\phi\in \Phi}\sum_{v\in V(\Phi)} (\alpha_Y f_{H\setminus e}(\phi_{e}))\\
  &\leq& F(\mathbf{x})+  \frac{1}{\kappa} \sum_{e\in E}y^\diamond_{e}   F_{\mathbf{x}\setminus e}(e) \label{line:3}
   \end{eqnarray}
The first two inequalities are due to the submodularity of $f$. The third inequality is due to $ \sum_{\phi\in \mathcal{U}} \beta_\phi=1$. The last inequality is due to the definition of degree of independence (Definition \ref{defin}).

We next provide a lower bound on the increased utility of $F(\mathbf{y}(t))$ during one round of Algorithm \ref{alg:greedy-peak}. To simplify the notation, we use $\overline{\mathbf{y}}$ to denote the optimal solution to \textbf{P1} in round $t$.
%     \begin{eqnarray}
%  &&\sum_{x\in X_t} \delta(1-\delta)^{m-1} [\omega_x-(1-(1-\delta)^{m-1})\Delta] \\ \label{eq:0}
%  &\geq&  \sum_{x\in O} \delta(1-\delta)^{m-1} [\omega_x-(1-(1-\delta)^{m-1})\Delta] \\
%  &\geq& \delta(1-\delta)^{m-1} (\sum_{x\in O}(\mathbb{E}[f_{R_i(p(t-1))}(x)]-(1-(1-\delta)^{m-1})\Delta))\\
%    &\geq& \delta(1-\delta)^{m-1} (\sum_{x\in O}(\mathbb{E}[f_{R(p(t-1))}(x)]-(1-(1-\delta)^{m-1})\Delta))\label{eq:1}\\
%  &\geq& \delta(1-\delta)^{m-1} (OPT-F(p(t-1))-m(1-(1-\delta)^{m-1})\Delta)\label{eq:2}
%   \end{eqnarray}
\begin{eqnarray}
&&F(\mathbf{y}(t+\delta))-F(\mathbf{y}(t)) \nonumber\\
&\geq& \sum_{e\in E}\delta \overline{y}_{e} \prod_{e'\neq e}(1-\delta \overline{y}_{e'}) F_{\mathbf{y}(t)}(e) \label{line3}\\
&\geq& \sum_{e\in E}\delta \overline{y}_{e} \prod_{e'\neq e}(1-\delta \overline{y}_{e'}) (1-y_e(t))F_{\mathbf{y}(t)\setminus e}(e)  \label{line4}\\
&\geq& \sum_{e\in E}\delta \overline{y}_{e} (1-\delta)^{m-1} (1-y_e(t))F_{\mathbf{y}(t)\setminus e}(e)\nonumber\\
&=&\delta(1-\delta)^{m-1}\sum_{e\in E} \overline{y}_{e}  (1-y_e(t))F_{\mathbf{y}(t)\setminus e}(e)\nonumber\\
%&\geq& \delta(1-\delta)^{m-1}\sum_{e\in E} \left(\overline{y}_{e} (1-y_e(t))F_{\mathbf{y}(t)\setminus e}(e)-(1- (1-\delta)^{m-1})\Delta \right) \nonumber\\
&\geq&\delta(1-\delta)^{m-1}\sum_{e\in E} \overline{y}_{e}  (1-t\delta)F_{\mathbf{y}(t)\setminus e}(e)\label{line55}\\
&=&(1-t\delta)\delta(1-\delta)^{m-1}\sum_{e\in E} \overline{y}_{e} F_{\mathbf{y}(t)\setminus e}(e) \nonumber\\
&\geq&(1-t\delta)\delta(1-\delta)^{m-1}(\sum_{e\in E} y^\diamond_{e} F_{\mathbf{y}(t)\setminus e}(e)-2m\delta f(\pi^\diamond))\label{line5}\\
%&\geq&\delta(1-\delta)^{m-1}\sum_{e\in E} y^\diamond_{e} (1-y_e(t))F_{\mathbf{y}(t)\setminus e}(e) \nonumber\\
&\geq& (1-t\delta)\delta(1-\delta)^{m-1} \left(\kappa(f(\pi^\diamond)-F(\mathbf{y}(t)))-2m\delta f(\pi^\diamond)\right) \label{line6}\\
&=&(1-t\delta)\delta(1-\delta)^{m-1} \kappa \left((1-\frac{2m\delta}{\kappa})f(\pi^\diamond)-F(\mathbf{y}(t))\right) \nonumber\\
&\geq& (1-t\delta)\delta(1-\delta m) \kappa \left((1-\frac{2m\delta}{\kappa}) f(\pi^\diamond)-F(\mathbf{y}(t))\right) \nonumber\\
&\geq& (1-t\delta)\delta \kappa \left((1-\frac{(\kappa+2)m\delta}{\kappa}) f(\pi^\diamond)-F(\mathbf{y}(t))\right) \label{line7}
%&\geq& \delta(1-m\delta)\sum_{e\in E} \overline{y}_{e}  f_{\mathbf{y}(t)\setminus \mathbb{E}_v}(e) \nonumber\\
%&=& \delta(1-\frac{1}{mn})\sum_{e\in E} \overline{y}_{e}  f_{\mathbf{y}(t)\setminus \mathbb{E}_v}(e)\label{line:3}
\end{eqnarray}

Inequality (\ref{line3}) is due to Lemma 3.3 in \citep{calinescu2011maximizing}. Inequality (\ref{line4}) is due to $F_{\mathbf{y}(t)}(e)=(1-y_e(t))F_{\mathbf{y}(t)\setminus e}(e)$. Inequality (\ref{line55}) is due to $\min_{e\in E}{(1-y_e(t))} \geq 1-t\delta$. Inequality (\ref{line5}) is due to $\overline{\mathbf{y}}$ is an optimal solution to \textbf{P1} and $\mathbf{y}^\diamond \in P_{\mathcal{I}}$ is a feasible solution to \textbf{P1}, and the rest proof is similar to the one of Lemma 3.2 in \citep{calinescu2011maximizing}). Inequality (\ref{line6}) is due to Inequality (\ref{line:3}).

%For any $t\leq \frac{1}{2}$, we have $y_e(t)\leq 1/2$ for all $e\in E$, thus, $\min_{e\in E}{(1-y_e(t))} \geq 1/2$. It follows from  (\ref{line7}) that $F(\mathbf{y}(t+\delta))-F(\mathbf{y}(t)) \geq  \frac{\delta \kappa}{2} \left((1-\frac{(\kappa+2)m\delta}{\kappa}) f(\pi^\diamond)-F(\mathbf{y}(t))\right)$ for any $t\leq \frac{1}{2}$.
By induction, we have $F(\mathbf{y}(1))\geq (1-e^{-\sum_{t=1}^{1/\delta}(1-t\delta)\delta \kappa})(1-\frac{(\kappa+2)m\delta}{\kappa}) f(\pi^\diamond)$. It follows that $F(\mathbf{y}(1))\geq (1-e^{-\frac{\kappa(1-\delta)}{2}})(1-\frac{(\kappa+2)m\delta}{\kappa}) f(\pi^\diamond)$. Finally, since $\delta=\frac{1}{9 m^2}$, we have $F(\mathbf{y}(1))\geq (1-e^{-\frac{\kappa}{2}+\frac{\kappa}{18m^2}})(1-\frac{\kappa+2}{3m\kappa}) f(\pi^\diamond)\geq (1-e^{-\frac{\kappa}{2}+\frac{\kappa}{18m^2}}-\frac{\kappa+2}{3m\kappa})f(\pi^\diamond)$.
%Thus, we have $f(\pi^\diamond)-\kappa m(1- (1-\delta)^{m-1})\Delta-F(p(1))\leq \frac{1}{e}(f(\pi^\diamond)-\kappa m(1- (1-\delta)^{m-1})\Delta)$, thus $F(p(1))\geq (1-\frac{1}{e})(f(\pi^\diamond)-\kappa m(1- (1-\delta)^{m-1})\Delta)=(1-\frac{1}{e})(f(\pi^\diamond)-\kappa m(1-(1-\frac{1}{n^2})^{m-1})\Delta)$.
%
% Let $\epsilon=m(1-(1-\frac{1}{n^2})^{m-1})$. Since $m\leq n$, we have $m(1-(1-\frac{1}{n^2})^{m-1})\leq n(1-(1-\frac{1}{n^2})^{n-1}$. Thus $\lim_{n\rightarrow\infty}\epsilon \leq \lim_{n\rightarrow\infty} n(1-(1-\frac{1}{n^2})^{n-1}\leq\lim_{n\rightarrow\infty} n(1-(1-\frac{1}{n^2})^{n}=\lim_{n\rightarrow\infty} n(1-(1+\frac{1}{n})^{n}(1-\frac{1}{n})^{n})=\lim_{n\rightarrow\infty} n-\lim_{n\rightarrow\infty} n(e\cdot\frac{1}{e})=0$.
$\Box$

As a byproduct of our main theorem, we have the following corollary.
\begin{corollary}
The adaptivity gap of SSMDI is  $\frac{1}{\alpha (1-e^{-\frac{\kappa}{2}+\frac{\kappa}{18m^2}}-\frac{\kappa+2}{3m\kappa})}$.
\end{corollary}

\section{Adaptivity Gap of SSMDI with Prefix-closed Constraints}
In this section, we analyze the adaptivity gap of SSMDI with  prefix-closed constraints. Notice that the class of prefix-closed constraints contains any
downward-closed/packing constraint \citep{gupta2017adaptivity}. To facilitate our study, we first introduce a new form of degree of independence.
\begin{definition}
\label{Def2}[A Second Form of Degree of Independence] The  degree of independence $\gamma(\mathcal{D})\in[0,1]$ of  a known prior probability distribution $\mathcal{D}$ is defined as follows
\begin{equation}
\gamma(\mathcal{D}):=\min_{ e\in E,  V\subseteq E\setminus \{e\}, \phi_V, \phi'_V\in \mathcal{U}_V}  \frac{\mathbb{E}_{\Phi_e}[f_{\phi_V\cup \phi'_V}(\Phi_e)]}{\mathbb{E}_{\Phi'_e}[f_{\phi_V\cup \phi'_V}(\Phi'_e)]}
\end{equation}
where $\Phi'_e$ follows the distribution of $\mathcal{D}_e(\phi'_V)$, $\Phi_e$ follows the distribution of $\mathcal{D}_e(\phi_V)$, and $f_{\phi_V\cup \phi'_V}(\cdot)=f(\phi_V\cup \phi'_V\cup \cdot)-f(\phi_V\cup \phi'_V)$.  If the numerator and denominator are both 0, the ratio is considered to be 1.
\end{definition}

Clearly, when all items are realized independently, $\gamma(\mathcal{D})=1$. We next present the main theorem. We omit $\mathcal{D}$ from $\gamma(\mathcal{D})$  if it is clear from the context.

\begin{theorem}
\label{them}
The adaptivity gap of SSMDI with prefix-closed constraints is $\frac{1+\gamma}{\gamma}$.
\end{theorem}

\emph{Proof:}  The basic idea of our proof is to show that given any adaptive policy, we can build a non-adaptive policy whose utility is at least $\frac{\gamma}{1+\gamma}$ times the utility of the given adaptive policy. The way we construct such a non-adaptive policy is inspired by the one proposed in \citep{DBLP:journals/corr/abs-1902-01461}. However, our setting is more complicated than theirs due to the dependency among items.

\begin{definition}[Virtual Non-adaptive Policy]
Given any adaptive policy $\pi$, we  define a non-adaptive policy $\sigma$ as follows: randomly draw
a realization $\phi$ according to $\mathcal{D}$, then run $\pi$ on $\phi$ virtually to pick a group of items.  
\end{definition}

Consider the first $|S|$ items $S$ picked by $\pi$, denote by $\mathrm{A}$ the realization of $S$ when picked by $\pi$ and we also use  $\mathrm{A}$ to denote the virtual realization when picked by $\sigma$. Let $\mathrm{B}$ be the true realization of $S$ when picked by $\sigma$.  Let  $\pi_\mathrm{A}$  denote the sub-decision tree $\pi$ goes to when $\Phi_S = \mathrm{A}$. Similarly, let $\sigma_\mathrm{A}$ denote the sub-decision tree\footnote{Any adaptive policy can be represented as a decision tree: Each node in the decision tree represents an item, we first
pick the root item and observe its state, and then choose a subtree to go to.} $\sigma$ goes to when $\Phi_S = \mathrm{A}$.  Assume the first item picked by $\pi_\mathrm{A}$ and $\sigma_{\mathrm{A}}$ is $v$. Denote by $\Phi_v$ the realization of $v$ when picked by $\sigma_{\mathrm{A}}$ and also the virtual realization when picked by $\sigma_\mathrm{A}$, while $\Phi'_v$ be the true realization of $v$ when picked by $\sigma_{\mathrm{A}}$. Thus $\Phi_v$ follows the distribution of  $\mathcal{D}_v(\mathrm{A})$ and $\Phi'_v$ follows the distribution of $\mathcal{D}_v(\mathrm{B})$.

Given two realizations $\mathrm{A}$ and $\mathrm{B}$ of $S$, we redefine the expected utility of a policy based on a new utility function $f_{\mathrm{A}\cup \mathrm{B}}(\cdot)=f(\mathrm{A}\cup \mathrm{B}\cup\cdot)-f(\mathrm{A}\cup \mathrm{B})$. Let $\mathcal{R}(\pi_{\mathrm{A}}, f_{\mathrm{A}\cup \mathrm{\mathrm{B}}}\mid \mathrm{A})=\sum_{\phi \in \mathcal{U}} \beta^{\mathrm{A}}_\phi f_{\mathrm{A}\cup \mathrm{B}}(\cup_{e\in E(\pi_\mathrm{A}, \phi)}\phi_e)$, where $\beta^{\mathrm{A}}_\phi$ is the probability that $\phi$ is realized conditioned on  $\Phi_S = \mathrm{A}$, denote the expected utility of  $\pi_{\mathrm{A}}$ conditioned on $\Phi_S = \mathrm{A}$. Similarly, we use $\mathcal{R}(\sigma_{\mathrm{A}}, f_{\mathrm{A}\cup  \mathrm{B}}\mid \mathrm{B})=\sum_{\phi \in \mathcal{U}} \beta^{\mathrm{B}}_\phi f_{\mathrm{A}\cup \mathrm{B}}(\cup_{e\in E(\sigma_\mathrm{A}, \phi)}\phi_e)$, where $\beta^{\mathrm{B}}_\phi$ is the probability that $\phi$ is realized conditioned on $\Phi_S = \mathrm{B}$,   to denote the expected utility of $\sigma_{\mathrm{A}}$ conditioned on $\Phi_S = \mathrm{B}$. To prove this theorem, it suffice to prove that when $S=\emptyset$, $\mathcal{R}(\sigma_{\emptyset}, f_{\emptyset}\mid \emptyset)\geq \frac{\gamma}{1+\gamma} \mathcal{R}(\pi_{\emptyset}, f_{\emptyset}\mid \emptyset)$. Next we prove a even stronger result: for any $S$ and its realizations $\mathrm{A}$ and $\mathrm{B}$,
\begin{equation}
\label{dq}
\mathcal{R}(\sigma_{\mathrm{A}}, f_{\mathrm{A}\cup  \mathrm{B}}\mid \mathrm{B})\geq \frac{\gamma}{1+\gamma} \mathcal{R}(\pi_{\mathrm{A}}, f_{\mathrm{A}\cup \mathrm{B}}\mid \mathrm{A})
\end{equation}
We prove (\ref{dq}) through induction on the size of $S$.  Assume the deepness of $\pi$ is $T$, i.e.,  $\pi$ picks at most $T$ items. The proof of the base case when $|S|=T-1$ is as follows. Note that when $|S|=T-1$, $\pi_{\mathrm{A}}$ (and also $\sigma_\mathrm{A}$) picks at most one item. To avoid trivial cases, assume $\pi_{\mathrm{A}}$ (and also $\sigma_\mathrm{A}$) picks $v$,  it follows that
\[\mathcal{R}(\sigma_{\mathrm{A}}, f_{\mathrm{A}\cup  \mathrm{B}}\mid \mathrm{B})=\mathbb{E}_{\Phi_v}[f_{\mathrm{A}\cup \mathrm{B}}(\Phi_v)]\]
\[\mathcal{R}(\pi_{\mathrm{A}}, f_{\mathrm{A}\cup \mathrm{B}}\mid \mathrm{A})=\mathbb{E}_{\Phi'_v}[f_{\mathrm{A}\cup \mathrm{B}}(\Phi_v)]\]
% The proof of the base case when $d=T$ is as follows: since $\pi$ only picks one item at level $T$, it follows that
%\[\mathcal{R}(\sigma_{\mathrm{A}}, f_{\mathrm{A}\cup \mathrm{B}}\mid \mathrm{B})=\mathbb{E}_{\Phi_v}[f_{\mathrm{A}\cup \mathrm{B}}(\Phi_v)\mid \mathrm{B}]\]
%\[\mathcal{R}(\pi_{\mathrm{A}}, f_{\mathrm{A}\cup \mathrm{B}}\mid \mathrm{A})=\mathbb{E}_{\Phi_v}[f_{\mathrm{A}\cup \mathrm{B}}(\Phi_v)\mid \mathrm{A}]\]
Because $\mathbb{E}_{\Phi_v}[f_{\mathrm{A}\cup \mathrm{B}}(\Phi_v)]\geq \gamma \mathbb{E}_{\Phi'_v}[f_{\mathrm{A}\cup \mathrm{B}}(\Phi'_v)]$ due to Definition \ref{Def2}, we have
\[\mathcal{R}(\sigma_{\mathrm{A}}, f_{\mathrm{A}\cup \mathrm{B}}\mid \mathrm{B})\geq \gamma \mathcal{R}(\pi_{\mathrm{A}}, f_{\mathrm{A}\cup \mathrm{B}}\mid \mathrm{A})\geq \frac{\gamma}{1+\gamma} \mathcal{R}(\pi_{\mathrm{A}}, f_{\mathrm{A}\cup \mathrm{B}}\mid \mathrm{A})\]
We next prove the induction step, assume the first item picked by $\pi_{\mathrm{A}}$ (and also $\sigma_\mathrm{A}$) is $v$. We first derive an uppder bound on  $\mathcal{R}(\pi_{\mathrm{A}}, f_{\mathrm{A}\cup \mathrm{B}}\mid \mathrm{A})$. % It is easy to verify that the expected utility of $\pi$
%\[\mathcal{R}(\pi, f^i)=\mathbb{E}_\mathrm{A}[f^i(\mathrm{A})+\mathcal{R}(\pi_\mathrm{A}, f^i_{\mathrm{A}})]\]
%\[\mathcal{R}(\sigma, f^i)=\mathbb{E}_{\mathrm{A},\mathrm{B}}[f^i(\mathrm{B})+\mathcal{R}(\sigma_{\mathrm{A}}, f^i_{\mathrm{B}})]\]
\begin{align}
\mathcal{R}(\pi_{\mathrm{A}}, f_{\mathrm{A}\cup \mathrm{B}}\mid \mathrm{A})&=\mathbb{E}_{\Phi_v}[f_{\mathrm{A}\cup \mathrm{B}}(\Phi_v)+ \mathcal{R}(\pi_{\mathrm{A}\cup\Phi_v }, f_{\mathrm{A}\cup \mathrm{B} \cup \Phi_v}\mid \mathrm{A}\cup \Phi_v)]\nonumber\\
&\leq \mathbb{E}_{\Phi_v,\Phi'_v}[f_{\mathrm{A}\cup \mathrm{B}}(\Phi_v\cup\Phi'_v)+\mathcal{R}(\pi_{\mathrm{A}\cup\Phi_v }, f^i_{\mathrm{A}\cup \mathrm{B}\cup \Phi_v\cup \Phi'_v}\mid \mathrm{A}\cup \Phi_v)] \label{eq:3}\\
&\leq \mathbb{E}_{\Phi_v,\Phi'_v}[f_{\mathrm{A}\cup \mathrm{B}}(\Phi_v)+f_{\mathrm{A}\cup \mathrm{B}}(\Phi'_v)+\mathcal{R}(\pi_{\mathrm{A}\cup\Phi_v }, f^i_{\mathrm{A}\cup \mathrm{B}\cup \Phi_v\cup \Phi'_v}\mid \mathrm{A}\cup \Phi_v)]\nonumber\\
&\leq  (1+\frac{1}{\gamma})\mathbb{E}_{\Phi'_v}[f_{\mathrm{A}\cup \mathrm{B}}(\Phi'_v)]+\mathbb{E}_{\Phi_v,\Phi'_v}[\mathcal{R}(\pi_{\mathrm{A}\cup\Phi_v }, f^i_{\mathrm{A}\cup \mathrm{B}\cup \Phi_v\cup \Phi'_v}\mid \mathrm{A}\cup \Phi_v)] \label{eq:2}
\end{align}
Inequality (\ref{eq:3}) is due to $f_{\mathrm{A}\cup \mathrm{B}}$ is submodular. Inequality (\ref{eq:2}) is due to $\gamma \mathbb{E}_{\Phi_v}[f_{\mathrm{A}\cup \mathrm{B}}(\Phi_v)]\leq \mathbb{E}_{\Phi'_v}[f_{\mathrm{A}\cup \mathrm{B}}(\Phi'_v)]$. Next we derive a lower bound on $\mathcal{R}(\sigma_{\mathrm{A}}, f_{\mathrm{A}\cup \mathrm{B}}\mid \mathrm{B})$.
%Inequality (\ref{eq:!}) is due to $f^i$ is $\epsilon$-nearly submodular, thus $f^i(\mathrm{A}\cup \mathrm{B})\leq \frac{1}{\epsilon}g^i(\mathrm{A}\cup \mathrm{B}) \leq \frac{1}{\epsilon}(g^i(\mathrm{A})+g^i(\mathrm{B}))\leq \frac{1}{\epsilon^2}(f^i(\mathrm{A})+f^i(\mathrm{B}))$. Inequality (\ref{eq:!}) is due to the fact that $\mathrm{A}$ and $\mathrm{B}$ are drawn from the same distribution, thus $\mathbb{E}_{\mathrm{A},\mathrm{B}}[f^i(\mathrm{A})]=\mathbb{E}_{\mathrm{A},\mathrm{B}}[f^i(\mathrm{B})]$.
\begin{align}
\mathcal{R}(\sigma_{\mathrm{A}}, f_{\mathrm{A}\cup \mathrm{B}}\mid \mathrm{B})&=\mathbb{E}_{\Phi_v,\Phi'_v}[f_{\mathrm{A}\cup \mathrm{B}}(\Phi'_v) + \mathcal{R}(\sigma_{\mathrm{A}\cup\Phi_v }, f_{\mathrm{A}\cup \mathrm{B} \cup \Phi'_v} \mid \mathrm{B}\cup \Phi'_v)]\nonumber\\
&\geq \mathbb{E}_{\Phi_v,\Phi'_v}[f_{\mathrm{A}\cup \mathrm{B}}(\Phi'_v) +\mathcal{R}(\sigma_{\mathrm{A}\cup\Phi_v }, f_{\mathrm{A}\cup \mathrm{B} \cup \Phi_v\cup \Phi'_v}\mid \mathrm{B}\cup \Phi'_v)]\nonumber\\
&= \mathbb{E}_{\Phi'_v}[f_{\mathrm{A}\cup \mathrm{B}}(\Phi'_v)]+ \mathbb{E}_{\Phi_v,\Phi'_v}[\mathcal{R}(\sigma_{\mathrm{A}\cup\Phi_v }, f_{\mathrm{A}\cup \mathrm{B} \cup \Phi_v\cup \Phi'_v}\mid \mathrm{B}\cup \Phi'_v)]~\nonumber\\
&\geq \mathbb{E}_{\Phi'_v}[f_{\mathrm{A}\cup \mathrm{B}}(\Phi'_v)]+ \frac{\gamma}{1+\gamma}\mathbb{E}_{\Phi_v,\Phi'_v}[\mathcal{R}(\sigma_{\mathrm{A}\cup\Phi_v }, f_{\mathrm{A}\cup \mathrm{B} \cup \Phi_v\cup \Phi'_v}\mid \mathrm{A}\cup \Phi_v)]\label{eq:4}
\end{align}
Inequalities (\ref{eq:2}) and (\ref{eq:4}) imply that $\mathcal{R}(\sigma_{\mathrm{A}}, f_{\mathrm{A}\cup  \mathrm{B}}\mid \mathrm{B})\geq \frac{\gamma}{1+\gamma} \mathcal{R}(\pi_{\mathrm{A}}, f_{\mathrm{A}\cup \mathrm{B}}\mid \mathrm{A})$.
%To prove inequality (\ref{eq:!}), it suffice to show that  $f^i_{\mathrm{A}\cup \mathrm{B}}$ is monotone and $\epsilon$-nearly submodular. \mathrm{B}ased on the definition of $f^i_{\mathrm{A}\cup \mathrm{B}}$, we have $f^i_{\mathrm{A}\cup \mathrm{B}}(S)=f^i(\mathrm{A}\cup \mathrm{B}\cup S)-f^i(\mathrm{A}\cup \mathrm{B})$. \mathrm{B}ecause $f^i$ is submodular, we have $\epsilon g^i(\mathrm{A}\cup \mathrm{B}\cup S)\leq f^i(\mathrm{A}\cup \mathrm{B}\cup S)\leq \frac{1}{\epsilon}g^i(\mathrm{A}\cup \mathrm{B}\cup S)$, it follows that $\epsilon g^i(\mathrm{A}\cup \mathrm{B}\cup S) - f^i(\mathrm{A}\cup \mathrm{B}) \leq f^i_{\mathrm{A}\cup \mathrm{B}}(S)\leq  \frac{1}{\epsilon}g^i(\mathrm{A}\cup \mathrm{B}\cup S)-f^i(\mathrm{A}\cup \mathrm{B})$. 
$\Box$

When $\gamma=1$, i.e., items are independent, Theorem \ref{them} reduces to the following corollary.
\begin{corollary}
\citep{DBLP:journals/corr/abs-1902-01461} The adaptivity gap is  $2$ when items are independent.
\end{corollary}
\section{Conclusion}
Previous studies on SSM often assume that items are independent, however, this assumption may not always hold. In this paper, we study SSM with dependent items.   To capture the impact of item dependency, we first introduce the concept of degree of independence. Then we propose a non-adaptive policy whose performance is close to the optimal adaptive policy. In particular, we prove that our non-adaptive policy achieves an approximation ratio whose value is depending on the degree of independence, the number of items, and the performance loss due to rounding. We also extend our study to the case with prefix-closed constraints and derive an adaptivity gap that depends on the (second form of) degree of independence.

%% The file named.bst is a bibliography style file for \mathrm{B}ibTeX 0.99c
\bibliographystyle{ormsv080}
\bibliography{reference}

\end{document}